# Econophysics of Asset Price, Return and Multiple Expectations

Victor Olkhov

TVEL, Kashirskoe sh. 49, Moscow, 115409, Russia

victor.olkhov@gmail.com

**Abstract**

This paper describes asset price and return disturbances as result of relations between transactions and multiple kinds of expectations. We show that disturbances of expectations can cause fluctuations of trade volume, price and return. We model price disturbances for transactions made under all types of expectations as weighted sum of partial price and trade volume disturbances for transactions made under separate kinds of expectations. Relations on price allow present return as weighted sum of partial return and trade volume "return" for transactions made under separate expectations. Dependence of price disturbances on trade volume disturbances as well as dependence of return on trade volume "return" cause dependence of volatility and statistical distributions of price and return on statistical properties of trade volume disturbances and trade volume "return" respectively.

Keywords: financial transactions; expectations; economic space; asset price; return

JEL: C02, C60, E32, F44, G00

---

This research did not receive any specific grant or financial support from TVEL or funding agencies in the public, commercial, or not-for-profit sectors.

## 1. Introduction

Asset pricing, price and return fluctuations and expectations are the key issues of modern finance. Publications on these problems account thousands and we refer (Campbell, 1985; Campbell and Cochrane, 1995; Heaton and Lucas, 2000; Cochrane, 2001; Cochrane and Cupl, 2003; Cochrane, 2017) as very short list with clear, precise and general treatment of the problem. Expectations as senior factors of finance and price dynamics are studied at least since Muth (1961) and by numerous further papers (Fama, 1965; Lucas, 1972; Sargent and Wallace, 1976; Hansen and Sargent, 1979; Brunnermeier and Parker, 2005; Dominitz and Manski, 2005; Greenwood and Shleifer, 2014; Lof, 2014; Manski, 2017). Description of return (Keim and Stambaugh, 1986; Mandelbrot, Fisher and Calvet, 1997; Fama, 1998; van Binsbergen and Koijen, 2017) and studies of statistical properties of markets, price and return (Brock and Hommes, 1998; Plerou et.al., 1999; Andersen et.al., 2001; Plerou et.al., 2001; Gabaix et.al., 2003; Stanley et.al., 2008; Hansen, 2013; Greenwald, Lettau and Ludvigson, 2014; Gontis et.al., 2016) present only small part of publications on these problems.

This paper develops our model of price and return disturbances induced by relations between transactions and expectations (Olkhov, 2018b). In brief our approach can be presented as follows. Any price and return movements of particular asset occur only after transactions with this asset are performed. Each transaction with particular asset defines trade volume and price of the deal and decision to perform this transaction is made under certain expectations. Thus relations between transactions and expectations define price and return dynamics and fluctuations. In this paper we study the model that describes impact of multiple kinds of expectations on transactions and cause price and return fluctuations with different frequencies. Transactions made under particular expectation define trade volume and price disturbances for this particular expectation. We describe price and return disturbances for transactions made under different expectations as function of price disturbances determined by transactions made under each particular expectations. Such representation uncovers relations between price disturbances determined by all possible kinds of expectations and trade volume disturbances defined by separate expectations. Price of transactions performed under separate expectations defines partial return for each expectation. We describe cumulative return under all possible kinds of expectations as function of partial return for each expectation.

Representation of price disturbances as weighted sum of partial price disturbances and trade volume disturbances determined by separate expectations impact statistical properties of price disturbances. The similar relations valid for return fluctuations. Statistical distributions of price and return fluctuations can depend on statistical properties of trade volume and on trade volume "return" (12.3) disturbances and should be studied further.

The rest of the paper unfolds as follows. In Sec.2 we present model setup and argue main definitions for transactions, trade value and volume, price, return and expectations. In Sec.3 we derive economic equations for transactions and in Sec.4 we introduce economic equations for expectations. In Sec. 5 we discuss simple model interactions between transactions and expectations and derive representation for disturbances of price as weighted sum of partial prices and trade volume disturbances. For return we obtain representation as weighted sum of partial return and partial trade "volume return". Conclusions are in Sec.6.

## 2. Model setup

Impact of expectations on price trends and fluctuations are studied at least since Muth (1961) and numerous papers (Lucas, 1972; Hansen and Sargent, 1979; Blume and Easley, 1984; Brunnermeier and Parker, 2005; Dominitz and Manski, 2005; Greenwood and Shleifer, 2014; Manski, 2017). In (Olkhov, 2018b) we present approximation that takes into account possible impact of expectations on transactions with particular assets performed by economic agents at Exchange. This approximation describes mutual interactions between transactions and expectations and models simple price fluctuations under action of expectations.

In current paper we describe possible impact of numerous kinds of expectations on price and return fluctuations.

Expectations are important characteristics of economic agents. Agents have economic and financial variables like assets and debts, credits and investment, consumption and labor and etc. Expectations of agents complement economic and financial properties of agents. Agents may have multiple expectations about future dynamics of any of economic or financial variables. Moreover, economic agents may establish their expectations about expectations of other agents and so on. Possible amount of expectations that can impact performance of economic or financial transactions exceed amount of economic variables and their "econometrics" increase complexity of economic modeling. Economic and financial transactions are only tools that change value of economic and financial variables of agents and thus define economic and financial evolution. In (Olkhov, 2018b) we assumed that each transaction is performed under single kind of expectations. For example Buy-Sell transactions of selected assets at Exchange can be performed under expectations of future returns. Meanwhile, different agents can use different kinds of expectations to take decisions on transactions. Moreover, even same agent can use different sorts of expectations to take decisions on Buy-Sell transactions of same asset. For example, Buy-Sell transactions can be performed under expectations of rate of inflation, currency exchange rate, GDP growth, weather forecasts, future returns and etc. Numerous kinds of expectations impact Buy-Sell transactions of the same assets. Thus various sorts of expectations those approve transactions of same assets can impact price and return. Below we model impact of multiple sorts of expectations on price and return fluctuations.

*2.1. Economic space*

Our approach is based on economic space notion. Below we present brief definitions and for details refer to (Olkhov, 2016a-2018b). Let's assume that all economic agents like banks, funds, companies, households and etc., can execute Buy-Sell transaction with particular assets at Exchange. There are a lot of risks that impact economic agents like credit risks, inflation risks, market risks and many other (Wilier, 1901; Horcher, 2005; McNeil, Frey and Embrechts, 2005; Skoglund and Chen, 2015). Let's treat all risks as factors that impact economic agents, their transactions and entire economics. For large banks and corporations risk assessments are provided by rating companies as Moody's, Fitch, S&P (Metz and Cantor, 2007; Chane-Kon, et.al, 2010; Kraemer and Vazza, 2012). Ratings take value of risk grades as *AAA, A, BB, C* and etc., and follow the risk rating methodologies (Staff U.S SEC, 2012; S&P, 2014; Pitman and Moss, 2016). We propose (Olkhov, 2016a-2018b) regard risk grades *AAA, A, BB, C* and etc., as points $x_1,...x_m$ of discrete economic space that imbed economic agents by their risk ratings $\boldsymbol{x}$. Ratings of single risk distribute agents over points of one-dimensional economic space. Assessments of two or more risks distribute agents by their risk ratings on economic space with dimension two or more. Let's assume that risk assessment methodology can utilize continuous risk grades. One can always assume that risk grades take value between most secure grade equals $0$ and most risky grade equals $X$. Assessments of agent's ratings for single risk fill interval $(0,X)$ of *economic domain* on space $R$. Assessments of agent's ratings for $n$ risks define agent's coordinates on economic domain on $R^n$. Let's propose that econometrics provide sufficient data to evaluate risk assessments for all economic agents of entire economics and there are $n=1,2,3$ major risks and their impact cause main action on agents and macroeconomics. Assessments of agents for $n$ major risks distribute them over economic domain on economic space $R^n$. We refer to (Olkhov, 2016a – 2018b) for additional details. For brevity let's further note economic space as e-space and economic agents as e-particles (economic particles). We use roman letters *f, t,* etc., to define scalar and bold letters $\boldsymbol{x}$, $\boldsymbol{B}$, $\boldsymbol{P}$, etc., to define vector variables.

*2.2. Transactions on e-space*

Let's model Buy-Sell transactions of e-particles at Exchange with particular assets. Let's assume that economics is described by $n$ major risks and each e-particle of the entire economics is determined by risk coordinates $\boldsymbol{x}$ on e-space $R^n$. Different e-particles may have same risk coordinates and we assume that at point $\boldsymbol{x}$ on e-space there are $N(t,\boldsymbol{x})$ different e-particles. Let's assume that each e-particle $i=1,2...N(t,\boldsymbol{x})$ at point $\boldsymbol{x}$ on e-space at moment $t$ under action of different types $k=1,2,...K$ of expectations $\boldsymbol{ex}_{ik}(t,\boldsymbol{x})$ perform transactions $\boldsymbol{tr}_{ik}(t,\boldsymbol{x})$ with particular assets. Transaction $\boldsymbol{tr}_{ik}(t,\boldsymbol{x})$ are performed under expectations of type $k$ between e-particle $i$ at point $\boldsymbol{x}$ and Exchange. Let's state that transactions $\boldsymbol{tr}_{ik}(t,\boldsymbol{x})$ define trading volumes $Q_{ik}(t,\boldsymbol{x})$ and trading values $SV_{ik}(t,\boldsymbol{x})$ (1.1):

$$\boldsymbol{tr_{ik}}(t,\boldsymbol{x}) = \left(Q_{ik}(t,\boldsymbol{x}); SV_{ik}(t,\boldsymbol{x})\right) \tag{1.1}$$

$$\boldsymbol{ex}_{ik}(t,\boldsymbol{x}) = \left(ex_{ikQ}(t,\boldsymbol{x});\ ex_{ikSV}(t,\boldsymbol{x})\right);\ k = 1,\dots K \tag{1.2}$$

We define transactions $\boldsymbol{tr}_{ik}(t,\boldsymbol{x})$ as two component functions (1.1) that are performed under action of expectations $\boldsymbol{ex}_{ik}(t,\boldsymbol{x})$ (1.2). Expectations $ex_{ikQ}$ are responsible for decisions on trading volumes $Q_{ik}(t,\boldsymbol{x})$ and expectations $ex_{ikQ}$ are responsible for decisions on trading values $SV_{ik}(t,\boldsymbol{x})$ of transactions $\boldsymbol{tr}_{ik}(t,\boldsymbol{x})$ (1.1). Thus different types $k=1,2...K$ of expectations approve different shares of transactions with same assets. Assets prices $p_{ik}(t,\boldsymbol{x})$ of transactions (1.1) are determined by obvious relations (1.3):

$$SV_{ik}(t,\boldsymbol{x}) = p_{ik}(t,\boldsymbol{x})Q_{ik}(t,\boldsymbol{x}) \tag{1.3}$$

Description of transactions performed by numerous agents under various kinds of expectations is a very complex problem. To simplify the model let's aggregate transactions and expectations of agents that have coordinates $\boldsymbol{x}$. Aggregation of transactions performed by e-particles at point $\boldsymbol{x}$ transfer description of transactions made by separate e-particles to description of transactions made by all agents at point $\boldsymbol{x}$. The same we do for description of expectations. We replace description of transactions and expectations of separate agents by modeling aggregated transactions and expectations made by e-particles at point $\boldsymbol{x}$. This transition average and rougher description of transactions and simplifies modeling expectations and transactions of entire economics. Such approximation has parallels to hydrodynamic approximation in physics of fluids and we shall note it as economic hydrodynamic-like approximation. Let's present above considerations in a more formal manner.

Let's remind (Olkhov, 2017d; 2018a; 2018b) that risk coordinates $\boldsymbol{x}=(x_1,...x_n)$ of e-particles on economic domain of $n$-dimensional e-space $R^n$ are reduced by min and max values:

$$0 \le x_i \le X_i\,, i = 1, \dots n \tag{1.4}$$

Here $x_i=0$ define most secure and $X_i$ define most risky grades for risk $i$. One can always set $X_i=1$ but we use notion $X_i$ for convenience. Relations (1.4) define economic domain and all economic agents (e-particles) have their risk coordinates inside economic domain (1.4). Let's assume that a unit volume $dV(\boldsymbol{x})$ at point $\boldsymbol{x}$ contains many e-particles but scales $dV_i$ (1.5) of a unit volume $dV(\boldsymbol{x})$ are small to compare with scales $X_i$ of economic domain (1.4):

$$dV_i \ll X_i\,, i = 1, \dots n \;\; ; \;\; dV = \prod_{i=1,..n} dV_i \tag{1.5}$$

Let's state that e-particles at moment $t$ have coordinates $\boldsymbol{x}=(x_1,...x_n)$ and velocities $\boldsymbol{v}=(v_1,...v_n)$. Velocities $\boldsymbol{v}=(v_1,...v_n)$ describe change of risk coordinates of e-particles. Let's define transaction $\boldsymbol{Tr}_k(t,\boldsymbol{x})$ as the sum of all transactions $\boldsymbol{tr}_{ik}(t,\boldsymbol{x})$ made under expectations $\boldsymbol{ex}_{ik}(t,\boldsymbol{x})$ of type $k$ from e-particles $i$ with coordinates $\boldsymbol{x}$ in a unit volume $dV(\boldsymbol{x})$ averaged during time $\Delta$:

$$\boldsymbol{Tr}_k(t,\boldsymbol{x}) = \big(Q_k(t,\boldsymbol{x}); SV_k(t,\boldsymbol{x})\big) = \sum_{i\in dV(\boldsymbol{x});\,\Delta} \boldsymbol{tr}_{ik}(t,\boldsymbol{x}) \tag{2.1}$$

$$Q_k(t,\boldsymbol{x}) = \sum_{i\in dV(\boldsymbol{x});\Delta} Q_{ik}(t,\boldsymbol{x}) \quad ; \quad SV_k(t,\boldsymbol{x}) = \sum_{i\in dV(\boldsymbol{x});\Delta} SV_{ik}(t,\boldsymbol{x})$$

$$\boldsymbol{tr}_{ik}(t,\boldsymbol{x}) = \big(Q_{ik}(t,\boldsymbol{x}); SV_{ik}(t,\boldsymbol{x})\big)$$

$$\sum_{i\in dV(\boldsymbol{x});\,\Delta} \boldsymbol{tr}_{ik}(t,\boldsymbol{x}) = \frac{1}{\Delta}\int_t^{t+\Delta} d\tau \, \sum_{i\in dV(\boldsymbol{x})} \boldsymbol{tr}_{ik}(\tau,\boldsymbol{x}) \tag{2.1.1}$$

Let's use $i \in dV(\boldsymbol{x})$ to denote that coordinates $\boldsymbol{x}$ of e-particle $i$ belong to a unit volume $dV(\boldsymbol{x})$. Let's use left hand sum (2.1.1) to denote averaging during time $\Delta$ in a unit volume $dV(\boldsymbol{x})$. Prices $p_{ik}(t,\boldsymbol{x})$ of transaction $\boldsymbol{tr}_{ik}(t,\boldsymbol{x})$ executed by e-particle $i$ at point $\boldsymbol{x}$ are determined by (1.3). Transaction $\boldsymbol{Tr}_k(t,\boldsymbol{x})$ as function of point $\boldsymbol{x}$ defines price $p_k(t,\boldsymbol{x})$ at point $\boldsymbol{x}$ (2.2) as

$$SV_k(t,x) = p_k(t,x)Q_k(t,x) \tag{2.2}$$

Aggregations by scales of unit volume $dV$ and averaging during time $\Delta$ move description of transactions of separate e-particles to description of transactions as function of $\boldsymbol{x}$ on e-space. Transactions $\boldsymbol{Tr}_k(t,\boldsymbol{x})$ become properties of points $\boldsymbol{x}$ of e-space but not properties of separate e-particles. Transactions $\boldsymbol{Tr}_k(t,\boldsymbol{x})$ determine transactions $\boldsymbol{Tr}(t,\boldsymbol{x})$ performed by all e-particles (agents) in a unit volume $dV$ under all possible expectations of all types $k=1,...K$ as:

$$\boldsymbol{Tr}(t,\boldsymbol{x}) = \sum_{k=1,..K} \boldsymbol{Tr}_k(t,\boldsymbol{x}) \;=\big(Q(t,\boldsymbol{x}); SV(t,\boldsymbol{x})\big) \tag{2.3}$$

$$Q(t,\boldsymbol{x}) = \sum\nolimits_{k=1,..K} Q_k(t,\boldsymbol{x}) \quad ; \quad SV(t,\boldsymbol{x}) = \sum\nolimits_{k=1,..K} SV_k(t,\boldsymbol{x})$$

Transactions $\boldsymbol{Tr}(t,\boldsymbol{x})$ taken under all expectation $k=1,..K$ define price $p(t,\boldsymbol{x})$ (2.4) as

$$SV(t,\boldsymbol{x}) = p(t,\boldsymbol{x})Q(t,\boldsymbol{x}) \tag{2.4}$$

To describe evolution of transactions $\boldsymbol{Tr}(t,\boldsymbol{x})$ one should model evolution of transactions $\boldsymbol{Tr}_k(t,\boldsymbol{x})$ made under expectations of type $k$. To describe evolution of transactions $\boldsymbol{Tr}_k(t,\boldsymbol{x})$ one should take into account (Olkhov, 2018b) motions of transactions $\boldsymbol{Tr}_k(t,\boldsymbol{x})$ that are induced by velocities $\boldsymbol{\upsilon}=(\upsilon_1,\dots\upsilon_n)$ of separate e-particles on e-space. Indeed, velocities $\boldsymbol{\upsilon}_i(t,\boldsymbol{x})=(\upsilon_1,\dots\upsilon_n)$ describe change of risk coordinates $\boldsymbol{x}$ of e-particle $i$ and thus describe *flows* of transactions $\boldsymbol{tr}_{ik}(t,\boldsymbol{x})$ of e-particle $i$ on e-space *alike to* flows of fluids. The product of velocity $\boldsymbol{\upsilon}_i(t,\boldsymbol{x})$ and transactions $\boldsymbol{tr}_{ik}(t,\boldsymbol{x})$ describe amount of transactions $\boldsymbol{tr}_{ik}(t,\boldsymbol{x})$ carried out in the direction of velocity $\boldsymbol{\upsilon}_i(t,\boldsymbol{x})$. Such a product $\boldsymbol{\upsilon}_i(t,\boldsymbol{x})\boldsymbol{tr}_{ik}(t,\boldsymbol{x})$ is *alike to "impulse"* of transaction. We use notion *"impulse"* to outline only one parallel between motion of transactions and motion of particles in physics: *"impulse"* $\boldsymbol{\upsilon}_i(t,\boldsymbol{x})\boldsymbol{tr}_{ik}(t,\boldsymbol{x})$ is an *additive* variable. Let's define transactions *"impulses"* $\boldsymbol{p}_{ik}$ of e-particle $i$ under expectations of type $k$ as (2.5.1):

$$\boldsymbol{p}_{ik}(t,\boldsymbol{x}) = (\boldsymbol{p}_{ikQ}(t,\boldsymbol{x}); \boldsymbol{p}_{ikSV}(t,\boldsymbol{x})) = \big(Q_{ik}(t,\boldsymbol{x})\,\boldsymbol{v}_i(t,\boldsymbol{x})\,;\, SV_{ik}(t,\boldsymbol{x})\,\boldsymbol{v}_i(t,\boldsymbol{x})\big) \tag{2.5.1}$$

Transactions "impulses" $\boldsymbol{p}_{ik}(t,\boldsymbol{x})$ are additive and sum (2.3.2) of "impulses" of two e-particles 1 and 2 $\boldsymbol{p}_{1k}(t,\boldsymbol{x}) + \boldsymbol{p}_{2k}(t,\boldsymbol{x})$ equals impulse $\boldsymbol{p}_k(t,\boldsymbol{x})$ of group of two e-particles:

$$\boldsymbol{p}_{\boldsymbol{k}}(t,\boldsymbol{x}) = \big(Q_k(t,\boldsymbol{x})\,\boldsymbol{v}_{kQ}(t,\boldsymbol{x})\,;\, SV_k(t,\boldsymbol{x})\,\boldsymbol{v}_{kSV}(t,\boldsymbol{x})\big) = \boldsymbol{p}_{1k}(t,\boldsymbol{x}) + \boldsymbol{p}_{2k}(t,\boldsymbol{x}) \tag{2.5.2}$$

$$Q_k(t,\boldsymbol{x})_k\,\boldsymbol{v}_{kQ}(t,\boldsymbol{x}) = Q_{1k}(t,\boldsymbol{x})\,\boldsymbol{v}_{\boldsymbol{1}}(t,\boldsymbol{x}) + Q_{2k}(t,\boldsymbol{x})\,\boldsymbol{v}_{\boldsymbol{2}}(t,\boldsymbol{x}) \tag{2.5.3}$$

$$SV_k(t,\boldsymbol{x})\,\boldsymbol{v}_{kSV}(t,\boldsymbol{x}) = SV_{1k}(t,\boldsymbol{x})\,\boldsymbol{v}_{\boldsymbol{1}}(t,\boldsymbol{x}) + SV_{2k}(t,\boldsymbol{x})\,\boldsymbol{v}_{\boldsymbol{2}}(t,\boldsymbol{x}) \tag{2.5.4}$$

Relations (2.5.1-2.5.4) show that risk velocities $\boldsymbol{\upsilon}_{kQ}(t,\boldsymbol{x})$ of the trade volumes $Q_k(t,\boldsymbol{x})$ and velocities $\boldsymbol{\upsilon}_{kSV}(t,\boldsymbol{x})$ of the transaction values $SV_k(t,\boldsymbol{x})$ can be different. Similar to (2.1) we aggregate transactions "impulses" under expectations of type $k$ of e-particles in a unit volume $dV(\boldsymbol{x})$ and average during time $\Delta$ and define transactions "impulses" and velocities under expectations of type $k=1,\dots K$ as functions of $\boldsymbol{x}$ (2.1.1):

$$\boldsymbol{P}_k(t,\boldsymbol{x}) = \big(\boldsymbol{P}_{kQ}(t,\boldsymbol{x})\,;\, \boldsymbol{P}_{kSV}(t,\boldsymbol{x})\big) = \sum\nolimits_{i\in dV(x);\Delta} \boldsymbol{p}_{ik}(t,\boldsymbol{x}) \tag{2.6.1}$$

$$\boldsymbol{P}_{kQ}(t,\boldsymbol{x}) = Q_k(t,\boldsymbol{x})\boldsymbol{v}_{kQ}(t,\boldsymbol{x}) = \sum\nolimits_{i\in dV(x);\Delta} Q_{ik}(t,\boldsymbol{x})\,\boldsymbol{v}_i(t,\boldsymbol{x}) \tag{2.6.2}$$

$$\boldsymbol{P}_{kSV}(t,\boldsymbol{x}) = SV_k(t,\boldsymbol{x})\boldsymbol{v}_{kSV}(t,\boldsymbol{x}) = \sum\nolimits_{i\in dV(x);\Delta} SV_{ik}(t,\boldsymbol{x})\,\boldsymbol{v}_i(t,\boldsymbol{x}) \tag{2.6.3}$$

$$\boldsymbol{p}_{ik}(t,\boldsymbol{x}) = (\boldsymbol{p}_{ikQ}(t,\boldsymbol{x}); \boldsymbol{p}_{ikSV}(t,\boldsymbol{x})) = \big(Q_{ik}(t,\boldsymbol{x})\boldsymbol{v}_i(t,\boldsymbol{x}); SV_{ik}(t,\boldsymbol{x})\boldsymbol{v}_i(t,\boldsymbol{x})\big) \tag{2.6.4}$$

$$\boldsymbol{v}_{\boldsymbol{k}}(t,\boldsymbol{x}) = (\boldsymbol{v}_{kQ}(t,\boldsymbol{x}); \boldsymbol{v}_{kSV}(t,\boldsymbol{x})) \tag{2.6.5}$$

Economic meaning of "impulses" is very simple. Impulses $\boldsymbol{p}_{ik}(t,\boldsymbol{x})$ describe flows of transactions under expectations of type $k$ of separate agents $i$ due to motion of agents on e-space. Impulses $\boldsymbol{P}_k(t,\boldsymbol{x})$ describe flow of transactions $\boldsymbol{Tr}_k(t,\boldsymbol{x})$ through a unit surface normal to velocity $\boldsymbol{v}_k(t,\boldsymbol{x})=(\boldsymbol{v}_{kQ}(t,\boldsymbol{x});\ \boldsymbol{v}_{kSV}(t,\boldsymbol{x}))$ (2.6.5) during time $dt$. Impulses $\boldsymbol{P}_k(t,\boldsymbol{x})$ describe flows of transactions under expectations of type $k$ induced by collective risk motion of all e-particles in a unit volume $dV(\boldsymbol{x})$ during time $\Delta$. Impulses $\boldsymbol{P}_k(t,\boldsymbol{x})$ of transactions under expectations of type $k$ define impulses $\boldsymbol{P}(t,\boldsymbol{x})$ (2.1-2.4; 2.7.1 - 2.7.3) of transactions $\boldsymbol{Tr}(t,\boldsymbol{x})$ performed under expectations of all types $k=1,..K$:

$$\boldsymbol{P}(t,\boldsymbol{x}) = \sum_{k=1,..K} \boldsymbol{P}_k(t,\boldsymbol{x}) = \left(\boldsymbol{P}_Q(t,\boldsymbol{x})\,;\ \boldsymbol{P}_{SV}(t,\boldsymbol{x})\right) \tag{2.7.1}$$

$$\boldsymbol{P}_Q(t,\boldsymbol{x}) = Q(t,\boldsymbol{x})\boldsymbol{v}_Q(t,\boldsymbol{x}) = \sum_{k=1,..K} Q_k(t,\boldsymbol{x})\boldsymbol{v}_{kQ}(t,\boldsymbol{x}) \tag{2.7.2}$$

$$\boldsymbol{P}_{SV}(t,\boldsymbol{x}) = SV(t,\boldsymbol{x})\boldsymbol{v}_{SV}(t,\boldsymbol{x}) = \sum_{k=1,..K} SV_k(t,\boldsymbol{x})\boldsymbol{v}_{SV}(t,\boldsymbol{x}) \tag{2.7.3}$$

Relations (2.1-2.7.3) define transactions $\boldsymbol{Tr}(t,\boldsymbol{x})$ performed under expectations of all types $k=1,..K$, and their "impulses" $\boldsymbol{P}(t,\boldsymbol{x})$ and velocities $\boldsymbol{v}(t,\boldsymbol{x})$ as functions of coordinates $\boldsymbol{x}$ on e-space. These relations (2.1-2.7.3) replace modeling transactions $\boldsymbol{tr}_{ik}(t,\boldsymbol{x})$ of separate e-particle $i$ at point $\boldsymbol{x}$ made under expectations of type $k$ by description of transactions $\boldsymbol{Tr}(t,\boldsymbol{x})$ made under all possible expectations with less accuracy on e-space determined by coarsening over a unit volume $dV$ and averaged during time $\Delta$. Such treatment has *certain parallels* to hydrodynamic approximation in physics (Landau and Lifshitz, 1987; Resibois and De Leener, 1977). Hydrodynamic approximation neglect granularity of separate particles and describes physical properties of the system as *continuous media*. We develop *similar* economic *continuous media* approximation to describe transactions of e-particles (agents) on e-space. Integral of transactions $\boldsymbol{Tr}(t,\boldsymbol{x})$ by variable $\boldsymbol{x}$ over e-space $R^n$ defines all transactions $\boldsymbol{Tr}(t)$ with particular assets performed by all e-particles in the entire economics at moment $t$.

*2.3. Expectations on e-space*

In this subsection let's argue expectations as functions on e-space. To do that let's underline that expectations should be treated as *intensive (non-additive)* variables. Indeed, expectations $\boldsymbol{ex}_{ik}(t,\boldsymbol{x})$ of e-particle (agent) $i$ of type $k$ those approve transactions $\boldsymbol{tr}_{ik}(t,\boldsymbol{x})$ have financial "weight" proportional to value of transactions $\boldsymbol{tr}_{ik}(t,\boldsymbol{x})$. Expectations $ex_{ikQ}(t,\boldsymbol{x})$ of trading volume $Q_{ik}(t,\boldsymbol{x})$ have "weight" proportional to trading volume $Q_{ik}(t,\boldsymbol{x})$. Expectations $ex_{ikSV}(t,\boldsymbol{x})$ of trading value $SV_{ik}(t,\boldsymbol{x})$ have "weight" proportional to trading value $SV_{ik}(t,\boldsymbol{x})$. It is reasonable to account expectations proportionally to amount of transactions performed under particular kind of expectations. To develop description of expectations on e-space and model mutual relations between transactions and expectations let's introduce *extensive* (*additive*) variables that we note as *expected transactions* $\boldsymbol{et}_{ik}(t,\boldsymbol{x})$ of type $k=1,..K,$ of e-particle $i$ :

$$\boldsymbol{et}_{ik}(t,\boldsymbol{x}) = \left(et_{ikQ}(t,\boldsymbol{x}); et_{ikSV}(t,\boldsymbol{x})\right) = \left(ex_{ikQ}(t,\boldsymbol{x})Q_{ik}(t,\boldsymbol{x}); ex_{ikSV}(t,\boldsymbol{x})SV_{ik}(t,\boldsymbol{x})\right) \quad (3.1)$$

Due to (3.1) sum of expected transactions $\boldsymbol{et}_{ik}(t,\boldsymbol{x})$ of group of $N$ e-particles can be presented as expected transaction $\boldsymbol{et}_{nk}(t,\boldsymbol{x})$ of entire group of $N$ e-particles. As example (3.2;3.3) present definition of expected transactions $et_{NkQ}(t,\boldsymbol{x})$ of type $k$ of trading volume $Q_{Nk}(t,\boldsymbol{x})$ for $N$ e-particles as follows:

$$et_{NkQ}(t,\boldsymbol{x}) = \sum_{i=1,..N} ex_{ik}(t,\boldsymbol{x})Q_{ik}(t,\boldsymbol{x}) = ex_{Nk}(t,\boldsymbol{x})Q_{Nk}(t,\boldsymbol{x}) \quad (3.2)$$

$$Q_{Nk}(t,\boldsymbol{x}) = \sum_{i=1,..N} Q_{ik}(t,\boldsymbol{x}) \quad (3.3)$$

Relations (3.2; 3.3) define expected transactions $et_{NkQ}(t,\boldsymbol{x})$ and expectations $ex_{NkQ}(t,\boldsymbol{x})$ of trading volume $Q_{Nk}(t,\boldsymbol{x})$ for group of $N$ e-particles. Similar to (3.2; 3.3) and (2.1; 2.3) let's define expected transactions $\boldsymbol{Et}_k(t,\boldsymbol{x})$ of type $k$ as sum of expected transactions $\boldsymbol{et}_{ik}(t,\boldsymbol{x})$ for all e-particles $i$ with coordinates $\boldsymbol{x}$ in a unit volume $dV(\boldsymbol{x})$ and averaged (2.1.1) during time $\Delta$:

$$\boldsymbol{Et}_k(t,\boldsymbol{x}) = \left(Et_{kQ}(t,\boldsymbol{x}); Et_{kSV}(t,\boldsymbol{x})\right) = \sum_{i\in dV(x);\,\Delta} \boldsymbol{et}_{ik}(t,\boldsymbol{x}) \quad (3.4)$$

$$Et_{kQ}(t,\boldsymbol{x}) = Ex_{kQ}(t,\boldsymbol{x})Q_k(t,\boldsymbol{x}) = \sum_{i\in dV(x);\Delta} et_{ikQ}(t,\boldsymbol{x})Q_{ik}(t,\boldsymbol{x}) \quad (3.5)$$

$$Et_{kSV}(t,\boldsymbol{x}) = Ex_{kSV}(t,\boldsymbol{x})SV_k(t,\boldsymbol{x}) = \sum_{i\in dV(x);\Delta} et_{ikSV}(t,\boldsymbol{x})SV_{ik}(t,\boldsymbol{x}) \quad (3.6)$$

Relations (3.4-3.6) and (2.5.3; 2.5.4) define aggregate expected transactions $\boldsymbol{Et}_k(t,\boldsymbol{x})$ and expectations $\boldsymbol{Ex}_k(t,\boldsymbol{x})$ (3.7) of type $k=1,..K$, of all e-particles in a unit volume $dV(\boldsymbol{x})$ averaged during time $\Delta$:

$$\boldsymbol{Ex}_k(t,\boldsymbol{x}) = \left(Ex_{kQ}(t,\boldsymbol{x}); Ex_{kSV}(t,\boldsymbol{x})\right) \quad (3.7)$$

Similar to (2.3) expected transactions $\boldsymbol{Et}_k(t,\boldsymbol{x})$ of type $k$ define expected transactions $\boldsymbol{Et}(t,\boldsymbol{x})$ and expectations $\boldsymbol{Ex}(t,\boldsymbol{x})$ for all types $k=1,..K$ of expectations in a unit volume $dV(\boldsymbol{x})$ at moment $t$ during time $\Delta$:

$$\boldsymbol{Et}(t,\boldsymbol{x}) = \sum_{k=1,..K} \boldsymbol{Et}_k(t,\boldsymbol{x}) = \left(Et_Q(t,\boldsymbol{x}); Et_{SV}(t,\boldsymbol{x})\right) \quad (3.8)$$

$$Et_Q(t,\boldsymbol{x}) = Ex_Q(t,\boldsymbol{x})Q(t,\boldsymbol{x}) = \sum_{k=1,..K} Ex_{kQ}(t,\boldsymbol{x})Q_k(t,\boldsymbol{x}) \quad (3.9)$$

$$Et_{SV}(t,\boldsymbol{x}) = Ex_{SV}(t,\boldsymbol{x})SV(t,\boldsymbol{x}) = \sum_{k=1,..K} Ex_{kSV}(t,\boldsymbol{x})SV_k(t,\boldsymbol{x}) \quad (3.10)$$

$$\boldsymbol{Ex}(t,\boldsymbol{x}) = \left(Ex_Q(t,\boldsymbol{x}); Ex_{SV}(t,\boldsymbol{x})\right) \quad (3.11)$$

Relations (3.8-3.10) define expected transactions $\boldsymbol{Et}(t,\boldsymbol{x})$ and expectations $\boldsymbol{Ex}(t,\boldsymbol{x})$ (3.9-3.11) as sum over all types of expectations $k=1,..K$. Evolution of expectations $\boldsymbol{Ex}(t,\boldsymbol{x})$ (3.9-3.11) is determined by evolution of expected transactions $\boldsymbol{Et}(t,\boldsymbol{x})$ (3.8-3.10). To describe evolution of *extensive* (additive) expected transactions $\boldsymbol{Et}(t,\boldsymbol{x})$ one should take into account motion of expected transactions $\boldsymbol{Et(t,x)}$ alike to motion of "*fluid*" induced by motion of separate e-particles on e-space. To do that let's introduce impulses $\boldsymbol{\Pi_k(t,x)}$ of expected transactions $\boldsymbol{Et_k(t,x)}$ of type k and impulses $\boldsymbol{\Pi(t,x)}$ of expected transactions $\boldsymbol{Et(t,x)}$ of all types $k=1,..K$ of expectations similar to definitions (2.6.1-2.7.3) of impulses $\boldsymbol{P(t,x)}$ of transactions $\boldsymbol{Tr(t,x)}$.

Indeed, expected transactions $\boldsymbol{et_{ik}(t,x)}$ of type $k$ for e-particle $i$ change their coordinates on e-space due to velocities $\boldsymbol{v}=(v_1,...v_n)$ of e-particles. Products of expected transaction $\boldsymbol{et_{ik}(t,x)}$ and velocity $\boldsymbol{v}=(v_1,...v_n)$ define (4.1.1-4.1.3) impulse $\boldsymbol{\eta_{ik}(t,x)}$ of expected transactions of type $k$ for e-particle $i$:

$$\boldsymbol{\eta}_{ik}(t,\boldsymbol{x}) = \left(\boldsymbol{\eta}_{ikQ}(t,\boldsymbol{x});\boldsymbol{\eta}_{ikSV}(t,\boldsymbol{x})\right) \tag{4.1.1}$$

$$\boldsymbol{\eta}_{ikQ}(t,\boldsymbol{x}) = ex_{ikQ}(t,\boldsymbol{x})Q_{ik}(t,\boldsymbol{x})\,\boldsymbol{v}_i(t,\boldsymbol{x}) \tag{4.1.2}$$

$$\boldsymbol{\eta}_{ikQ}(t,\boldsymbol{x}) = ex_{ikSV}(t,\boldsymbol{x})SV_{ik}(t,\boldsymbol{x})\,\boldsymbol{v}_i(t,\boldsymbol{x}) \tag{4.1.3}$$

$\boldsymbol{\eta_{ik}(t,x)}$ – describe "impulses" of expectations of trading volume $Q_{ik}(t,\boldsymbol{x})$ and $\boldsymbol{\eta_{ikSV}(t,x)}$ – describe "impulses" of expectations of trading value $SV_{ik}(t,\boldsymbol{x})$. "Impulse" $\boldsymbol{\eta_{ikQ}(t,x)}$ describes flow of expected transaction (3.1) due to motion of e-particle $i$ with velocity $\boldsymbol{v}_i(t,\boldsymbol{x})$. Aggregation of "impulses" $\boldsymbol{\eta_{ik}(t,x)}$ of type $k$ of all e-particles in a unit volume $dV(\boldsymbol{x})$ and averaging (2.1.1) during time $\Delta$ determines impulses of expectations of type $k$ at point $\boldsymbol{x}$ at moment $t$ similar to (2.6.1-6.6.5):

$$\boldsymbol{\Pi}_k(t,\boldsymbol{x}) = \left(\boldsymbol{\Pi}_{kQ}(t,\boldsymbol{x})\,;\,\boldsymbol{\Pi}_{kSV}(t,\boldsymbol{x})\right) = \sum_{i\in dV(x);\Delta}\boldsymbol{\eta}_{ik}(t,\boldsymbol{x}) \tag{4.2.1}$$

$$\boldsymbol{\Pi}_{kQ}(t,\boldsymbol{x}) = Et_{kQ}(t,\boldsymbol{x})\boldsymbol{v}_{keQ}(t,\boldsymbol{x}) = \sum_{i\in dV(x);\Delta} ex_{ikQ}(t,\boldsymbol{x})Q_{ik}(t,\boldsymbol{x})\,\boldsymbol{v}_i(t,\boldsymbol{x}) \tag{4.2.2}$$

$$\boldsymbol{\Pi}_{kSV}(t,\boldsymbol{x}) = Et_{kSV}(t,\boldsymbol{x})\boldsymbol{v}_{keSV}(t,\boldsymbol{x}) = \sum_{i\in dV(x);\Delta} ex_{ikSV}(t,\boldsymbol{x})SV_{ik}(t,\boldsymbol{x})\,\boldsymbol{v}_i(t,\boldsymbol{x}) \tag{4.2.3}$$

$$\boldsymbol{v}_{ke}(t,\boldsymbol{x}) = (\boldsymbol{v}_{keQ}(t,\boldsymbol{x});\boldsymbol{v}_{kSeV}(t,\boldsymbol{x})) \tag{4.2.4}$$

Due to (3.5; 3.6) relations (4.2.2; 4.2.3) take form:

$$\boldsymbol{\Pi}_{kQ}(t,\boldsymbol{x}) = Ex_{kQ}(t,\boldsymbol{x})Q_k(t,\boldsymbol{x})\boldsymbol{v}_{keQ}(t,\boldsymbol{x}) \tag{4.2.5}$$

$$\boldsymbol{\Pi}_{kSV}(t,\boldsymbol{x}) = Ex_{kSV}(t,\boldsymbol{x})SV_k(t,\boldsymbol{x})\boldsymbol{v}_{keSV}(t,\boldsymbol{x}) \tag{4.2.6}$$

Economic meaning of "impulses" $\boldsymbol{\Pi_k(t,x)}$ of expected transactions $\boldsymbol{Et_k(t,x)}$ of type $k$ is similar to meaning of "impulses" $\boldsymbol{P_k(t,x)}$ of transactions $\boldsymbol{Tr_k(t,x)}$: $\boldsymbol{\Pi_k(t,x)}$ describe flows of expected transactions $\boldsymbol{Et_k(t,x)}$ through a unit surface normal to velocity $\boldsymbol{v_{ke}(t,x)=(ve_{kQ}(t,x);\ v_{keSV}(t,x))}$ during time $dt$. Such a flow is induced by velocities $\boldsymbol{v}=(v_1,...v_n)$ of e-particles on e-space. "Impulses" $\boldsymbol{\Pi_k(t,x)}$ for expectations of type $k$ define "impulses" $\boldsymbol{\Pi(t,x)}$ for expectations of all types $k=1,..K$ similar to (2.7.1-2.7.3) as:

$$\boldsymbol{\Pi}(t,\boldsymbol{x}) = \sum_{k=1,..K}\boldsymbol{\Pi}_k(t,\boldsymbol{x}) \;= \left(\boldsymbol{\Pi}_Q(t,\boldsymbol{x})\,;\,\boldsymbol{\Pi}_{SV}(t,\boldsymbol{x})\right) \tag{4.3.1}$$

$$\boldsymbol{\Pi}_Q(t,\boldsymbol{x}) = Et_Q(t,\boldsymbol{x})\boldsymbol{v}_{eQ}(t,\boldsymbol{x}) = \sum_{k=1,..K} Et_{kQ}(t,\boldsymbol{x})\boldsymbol{v}_{keQ}(t,\boldsymbol{x}) \tag{4.3.2}$$

$$\boldsymbol{\Pi}_{SV}(t,\boldsymbol{x}) = Et_{SV}(t,\boldsymbol{x})\boldsymbol{v}_{eSV}(t,\boldsymbol{x}) = \sum_{k=1,..K} Et_{kSV}(t,\boldsymbol{x})\boldsymbol{v}_{keSV}(t,\boldsymbol{x}) \tag{4.3.3}$$

Relations (4.3.1-4.3.3) complete set on notions that are required for modeling evolution and mutual interaction between transactions $\boldsymbol{Tr}(t,\boldsymbol{x})$ and expectations $\boldsymbol{Ex}(t,\boldsymbol{x})$ on e-space. In the next Section we argue economic equations that model dynamics and interactions between transactions $\boldsymbol{Tr}(t,\boldsymbol{x})$ and expectations $\boldsymbol{Ex}(t,\boldsymbol{x})$.

## 3. Economic equations on transactions

Evolution of transactions $\boldsymbol{Tr}(t,\boldsymbol{x})$ (2.3) is determined by dynamics of transactions $\boldsymbol{Tr}_k(t,\boldsymbol{x})$ (2.1) taken under expectations $\boldsymbol{Ex}_k(t,\boldsymbol{x})$ (3.4-3.7). Let's follow (Olkhov, 2018a; 2018b) and derive economic equations that describe dynamics of transactions $\boldsymbol{Tr}_k(t,\boldsymbol{x})$ (2.1). Components $Q_k(t,\boldsymbol{x})$ and $SV_k(t,\boldsymbol{x})$ (5.1) of transactions $\boldsymbol{Tr}_k(t,\boldsymbol{x})$ :

$$\boldsymbol{Tr}_k(t,\boldsymbol{x}) = \big(Q_k(t,\boldsymbol{x}); SV_k(t,\boldsymbol{x})\big) \tag{5.1}$$

are extensive (additive) variables. There are two factors that change $Q_k(t,\boldsymbol{x})$ in a unit volume $dV$ at point $\boldsymbol{x}$ and moment $t$. The first factor describes change of $Q_k(t,\boldsymbol{x})$ in time as $\partial Q_k(t,\boldsymbol{x})/\partial t$. The second factor change value of extensive variable $Q_k(t,\boldsymbol{x})$ in a unit volume $dV$ due to flux of $Q_k(t,\boldsymbol{x})\boldsymbol{v}_{kQ}(t,\boldsymbol{x})$ through surface of a unit volume. Indeed, some value of $Q_k(t,\boldsymbol{x})$ can flow *out* or flow *in* a unit volume $dV$ during time $dt$ and that will change amount of $Q_k(t,\boldsymbol{x})$ in $dV$. Amount of flow $Q_k(t,\boldsymbol{x})$ is described by $Q_k(t,\boldsymbol{x})\boldsymbol{v}_{kQ}(t,\boldsymbol{x})$. Origin of such flux $Q_k(t,\boldsymbol{x})\boldsymbol{v}_{kQ}(t,\boldsymbol{x})$ is the motion of e-particles on e-space. Risk ratings of agents can change during time $dt$ and that is described by motion of e-particles on e-space. Velocities $\boldsymbol{v}_{ik}(t,\boldsymbol{x})$ of particular e-particle $i$ on e-space carry trading volumes $Q_{ik}(t,\boldsymbol{x})$ of this e-particle. Aggregates of motion of e-particles at point $\boldsymbol{x}$ define "impulses" $\boldsymbol{P}_k(t,\boldsymbol{x})$ (2.6.1-2.6.5) that describe fluxes of trading volume $Q_k(t,\boldsymbol{x})$, values $SV_k(t,\boldsymbol{x})$ and (4.2.1-4.2.6) describe fluxes of expected transactions $Et_{kQ}(t,\boldsymbol{x})$ and $Et_{kSV}(t,\boldsymbol{x})$. Flux of $Q_k(t,\boldsymbol{x})\boldsymbol{v}_{kQ}(t,\boldsymbol{x})$ through surface of a unit volume $dV$ can increase or decrease amount of trading volume $Q_k(t,\boldsymbol{x})$ in a unit volume $dV$. Due to divergence theorem (Strauss 2008, p.179) integral of flux $Q_k(t,\boldsymbol{x})\boldsymbol{v}_{kQ}(t,\boldsymbol{x})$ over surface of a unit volume $dV$ equals volume integral of divergence:

$$\oint ds\, Q_k(t,\boldsymbol{x})\, \boldsymbol{v}_{kQ}(t,\boldsymbol{x}) = \int dV\ \nabla\cdot\big(Q_k(t,\boldsymbol{x})\, \boldsymbol{v}_{kQ}(t,\boldsymbol{x})\big) \tag{5.2.1}$$

Thus total change of trading volume $Q_k(t,\boldsymbol{x})$ in a unit $dV$ is described as:

$$\int dV\, \Big[\frac{\partial}{\partial t} Q_k(t,\boldsymbol{x}) + \nabla\cdot\Big(Q_k(t,\boldsymbol{x})\, \boldsymbol{v}_{kQ}(t,\boldsymbol{x})\Big)\Big] \tag{5.2.2}$$

Hence we obtain change of trading volume $Q_k(t,\boldsymbol{x})$ in a unit volume $dV$ at point $\boldsymbol{x}$ as:

$$\frac{\partial}{\partial t} Q_k(t,\boldsymbol{x}) + \nabla\cdot\Big(Q_k(t,\boldsymbol{x})\, \boldsymbol{v}_{kQ}(t,\boldsymbol{x})\Big) \tag{5.2.3}$$

Due to (2.6.2):

$$\boldsymbol{P}_{kQ}(t,\boldsymbol{x}) = Q_k(t,\boldsymbol{x})\ \boldsymbol{v}_{kQ}(t,\boldsymbol{x})$$

and relations (5.2.3) can take form:

$$\frac{\partial}{\partial t} Q_k(t,\boldsymbol{x}) + \nabla \cdot \boldsymbol{P}_{kQ}(t,\boldsymbol{x}) \quad (5.2.4)$$

Relations (5.2.1-5.2.4) are reproduced in any treatise on physics of *continuous media* at least for fifty years (Batchelor, 1967; Resibois and De Leener, 1977; Landau and Lifshitz, 1987) and are valid for any extensive (additive) economic or financial characteristics defined similar to (2.1; 2.6.1-2.6.5; 3.4-3.6; 4.2.1-4.2.5) as aggregates of corresponding characteristics of e-particles (agents) at point $\boldsymbol{x}$ on e-space. Dynamics of any *continuous media* – with economic, financial or physical properties – follows relations (5.2.1-5.2.4). Let's underline important property of (5.2.1-5.2.4): integral in the right hand side of (5.2.1) over economic domain (1.4) equals integral of fluxes through surface outside of economic domain (1.4). Due to definition (1.4) of economic domain there are no agents and hence no economic or financial variables outside of economic domain. Thus integral (5.2.1) over economic domain (1.4) equals zero for any extensive economic variable or economic transactions. Hence integral for (5.2.4) over economic domain (1.4) equals simple ordinary time derivative for $Q_k(t)$:

$$\int dV\ \left[\frac{\partial}{\partial t} Q_k(t,\boldsymbol{x}) + \nabla \cdot \boldsymbol{P}_{kQ}(t,\boldsymbol{x})\right] = \frac{d}{dt} Q_k(t) \quad (5.2.5)$$

Equations (5.2.5) determine relations between equations (5.2.3; 5.2.4) and ordinary time derivatives of economic or financial variables of entire economics defined as functions of time *t* only. On e-space usual ordinary time derivative operator *d/dt* is replaced by relations (5.2.3) that take into account change of $Q_k(t,\boldsymbol{x})$ in a unit volume *dV* due to flux $Q_k(t,\boldsymbol{x})\boldsymbol{v}_{kQ}(t,\boldsymbol{x})$ induced by motion of e-particles on e-space caused by change of their risk ratings. Economic equations on trading volume $Q_k(t,\boldsymbol{x})$ describe balance between change of $Q_k(t,\boldsymbol{x})$ (5.2.3; 5.2.4) at point $\boldsymbol{x}$ and economic and financial factors that impact such a change. Let's note these factors as $\boldsymbol{F}_{kQ}(t,\boldsymbol{x})$. Transactions with trading volume $Q_k(t,\boldsymbol{x})$ may depend on trading value $SV_k(t,\boldsymbol{x})$ (2.1; 2.2), impulses $\boldsymbol{P}_k(t,\boldsymbol{x})$ (2.6.1-2.6.5) and on other economic or financial variables or other transactions performed with other assets or on other economic variables. As well decisions on transactions $\boldsymbol{Tr}_k(t,\boldsymbol{x})$ (2.1) are made under expectations $\boldsymbol{Ex}_k(t,\boldsymbol{x})$ (3.4).

To model impact of expectations $\boldsymbol{Ex}_k(t,\boldsymbol{x})$ on transactions $\boldsymbol{Tr}_k(t,\boldsymbol{x})$ let's propose that factors $\boldsymbol{F}_k(t,\boldsymbol{x})$ depend on expected transactions $\boldsymbol{Et}_k(t,\boldsymbol{x})$ (3.4) or their impulses $\boldsymbol{\Pi}_k(t,\boldsymbol{x})$ (4.2.1-4.2.6). Let's take equations on $Q_k(t,\boldsymbol{x})$ as:

$$\frac{\partial}{\partial t} Q_k(t,\boldsymbol{x}) + \nabla \cdot \left(Q_k(t,\boldsymbol{x})\ \boldsymbol{v}_{kQ}(t,\boldsymbol{x})\right) = F_{kQ}(t,\boldsymbol{x}) \quad (6.1)$$

The same considerations allows take equations on asset value $SV_k(t,\boldsymbol{x})$ ) (2.1; 2.2) as:

$$\frac{\partial}{\partial t}SV_k(t,\boldsymbol{x}) + \nabla\cdot\big(SV_k(t,\boldsymbol{x})\,\boldsymbol{v}_{kSV}(t,\boldsymbol{x})\big) = F_{kSV}(t,\boldsymbol{x}) \tag{6.2}$$

Relations (6.1; 6.2) permit take equations of transactions $\boldsymbol{Tr}_k(t,\boldsymbol{x})$ as:

$$\frac{\partial}{\partial t}\boldsymbol{Tr}_k(t,\boldsymbol{x}) + \nabla\cdot\big(\boldsymbol{Tr}_k(t,\boldsymbol{x})\,\boldsymbol{v}_k(t,\boldsymbol{x})\big) = \boldsymbol{F}_k(t,\boldsymbol{x}) \tag{6.3}$$

$$\boldsymbol{F}_k(t,\boldsymbol{x}) = \Big(F_{kQ}(t,\boldsymbol{x});\; F_{kSV}(t,\boldsymbol{x})\Big) \tag{6.4}$$

Due to relations (5.2.5) integrals of equations (6.3) over economic domain (1.4) give:

$$\frac{d}{dt}\boldsymbol{Tr}_k(t) = \int dV\left[\frac{\partial}{\partial t}\boldsymbol{Tr}_k(t,\boldsymbol{x}) + \nabla\cdot\big(\boldsymbol{Tr}_k(t,\boldsymbol{x})\,\boldsymbol{v}_k(t,\boldsymbol{x})\big)\right] = \int dV\,\boldsymbol{F}_k(t,\boldsymbol{x}) = \boldsymbol{F}_k(t) \tag{6.5}$$

Ordinary differential equations (6.5) describe evolution of cumulative transactions of all agents of entire economics with selected assets. Equations (6.1-6.4) depend on velocity $\boldsymbol{v}_k = (\boldsymbol{v}_{kQ};\boldsymbol{v}_{kSV})$ and hence economic equations that describe evolution of transactions in a closed form should incorporate equations on velocities $\boldsymbol{v}_k$ or impulses $\boldsymbol{P}_{kQ}(t,\boldsymbol{x})$ and $\boldsymbol{P}_{kSV}(t,\boldsymbol{x})$ (2.6.1-2.6.5). All reasons that ground relations (5.2.1-5.2.4) for trading volume $Q_k(t,\boldsymbol{x})$ are valid for any additive variables and impulses $\boldsymbol{P}_{kQ}(t,\boldsymbol{x})$ (2.6.1-2.6.5) also. All components of $\boldsymbol{P}_{kjQ}(t,\boldsymbol{x})$, $j=1,..n$ on $n$-dimensional e-space $R^n$ change in a unit volume due to change in time and due to flux of components $P_{kjQ}(t,\boldsymbol{x})$ through surface of a unit volume. Thus each components $P_{kjQ}(t,\boldsymbol{x})$ follow equations similar to (6.1) as:

$$\frac{\partial}{\partial t}P_{kjQ}(t,\boldsymbol{x}) + \nabla\cdot\Big(P_{kjQ}(t,\boldsymbol{x})\,\boldsymbol{v}_{kQ}(t,\boldsymbol{x})\Big) = G_{kjQ}(t,\boldsymbol{x})\,;\;\; j = 1,..n \tag{7.1}$$

or for impulses $\boldsymbol{P}_k(t,\boldsymbol{x})$:

$$\frac{\partial}{\partial t}\boldsymbol{P}_k(t,\boldsymbol{x}) + \nabla\cdot\Big(\boldsymbol{P}_k(t,\boldsymbol{x})\,\boldsymbol{v}_{kQ}(t,\boldsymbol{x})\Big) = \boldsymbol{G}_k(t,\boldsymbol{x}) \tag{7.2}$$

$$\boldsymbol{P}_k(t,\boldsymbol{x}) = \Big(\boldsymbol{P}_{kQ}(t,\boldsymbol{x})\boldsymbol{P}_{kSV}(t,\boldsymbol{x})\Big);\;\; \boldsymbol{G}_k(t,\boldsymbol{x}) = (\boldsymbol{G}_{kQ}(t,\boldsymbol{x})\boldsymbol{G}_{kSV}(t,\boldsymbol{x})) \tag{7.3}$$

$$\nabla\cdot\Big(\boldsymbol{P}_{kQ}(t,\boldsymbol{x})\,\boldsymbol{v}_{kQ}(t,\boldsymbol{x})\Big) = \sum_{i=1,\ldots n}\frac{\partial}{\partial x_i}(\,\boldsymbol{P}_{kQ}(t,\boldsymbol{x})\,\boldsymbol{v}_{kiQ}(t,\boldsymbol{x})) \tag{7.4}$$

Factors $\boldsymbol{G}_k(t,\boldsymbol{x})$ in the right side of (7.2) describe impact of economic and financial variables, their impulses, expectations or other transactions on evolution of impulses $\boldsymbol{P}_k(t,\boldsymbol{x})$. Economic equations (6.1-6.4) and (7.1-7.4) describe evolution of transactions $\boldsymbol{Tr}_k(t,\boldsymbol{x})$ and their impulses $\boldsymbol{P}_k(t,\boldsymbol{x})$ under action of economic and financial variables, expectations and other transactions determined by factors $\boldsymbol{F}_k(t,\boldsymbol{x})$ and $\boldsymbol{G}_k(t,\boldsymbol{x})$.

## 4. Economic equations on expectations

To describe mutual action of transactions and expectations let's derive economic equations on expectations. In Sec. 2.3 we argue that description of expectations should be developed via modeling extensive (additive) variables that we note as expected transactions $\boldsymbol{Et}_k(t,\boldsymbol{x})$ (3.4-3.6) and their impulses $\boldsymbol{\Pi}_k(t,\boldsymbol{x})$ (4.2.1-4.2.6). Evolution of additive variables like expected transactions $\boldsymbol{Et}_k(t,\boldsymbol{x})$ and their impulses $\boldsymbol{\Pi}_k(t,\boldsymbol{x})$ follows economic equations similar to (6.1-6.4) and (7.1-7.4). Economic equations on expected transactions $\boldsymbol{Et}_k(t,\boldsymbol{x})$ take form similar to equations on transactions $\boldsymbol{Tr}_k(t,\boldsymbol{x})$ (6.1-6.4):

$$\frac{\partial}{\partial t}\boldsymbol{Et}_k(t,\boldsymbol{x}) + \nabla\cdot\left(\boldsymbol{Et}_k(t,\boldsymbol{x})\,\boldsymbol{u}_k(t,\boldsymbol{x})\right) = \boldsymbol{Fe}_k(t,\boldsymbol{x}) \tag{8.1}$$

$$\boldsymbol{Et}_k(t,\boldsymbol{x}) = \left(Et_{kQ}(t,\boldsymbol{x}); Et_{kSV}\right)\ ;\quad \boldsymbol{Fe}_k(t,\boldsymbol{x}) = \left(Fe_{kQ}(t,\boldsymbol{x});\ Fe_{kSV}(t,\boldsymbol{x})\right) \tag{8.2}$$

$$\frac{\partial}{\partial t}Et_{kQ}(t,\boldsymbol{x}) + \nabla\cdot\left(Et_{kQ}(t,\boldsymbol{x})\,\boldsymbol{u}_{kQ}(t,\boldsymbol{x})\right) = Fe_{kQ}(t,\boldsymbol{x}) \tag{8.3}$$

$$\frac{\partial}{\partial t}Et_{kSV}(t,\boldsymbol{x}) + \nabla\cdot\left(Et_{kSV}(t,\boldsymbol{x})\,\boldsymbol{u}_{kSV}(t,\boldsymbol{x})\right) = Fe_{kSV}(t,\boldsymbol{x}) \tag{8.4}$$

Factors $\boldsymbol{Fe}_k(t,\boldsymbol{x})$ in the right hand side of (8.1 -8.4) describe action of economic and financial variables, transactions or their impulses and expected transactions on evolution of $\boldsymbol{Et}_k(t,\boldsymbol{x})$. To describe evolution of velocity $\boldsymbol{u}_k(t,\boldsymbol{x})=(\boldsymbol{u}_{kQ}(t,\boldsymbol{x});\ \boldsymbol{u}_{kSV}(t,\boldsymbol{x}))$ of expected transactions let's take equations on their impulses $\boldsymbol{\Pi}_k(t,\boldsymbol{x})$ (4.2.1-4.2.6) similar to (7.1-7.4):

$$\frac{\partial}{\partial t}\boldsymbol{\Pi}_k(t,\boldsymbol{x}) + \nabla\cdot\left(\boldsymbol{\Pi}_k(t,\boldsymbol{x})\,\boldsymbol{u}_{kQ}(t,\boldsymbol{x})\right) = \boldsymbol{Ge}_k(t,\boldsymbol{x}) \tag{8.5}$$

$$\frac{\partial}{\partial t}\boldsymbol{\Pi}_{kQ}(t,\boldsymbol{x}) + \nabla\cdot\left(\boldsymbol{\Pi}_{kQ}(t,\boldsymbol{x})\,\boldsymbol{u}_{kQ}(t,\boldsymbol{x})\right) = \boldsymbol{Ge}_{kQ}(t,\boldsymbol{x}) \tag{8.6}$$

$$\frac{\partial}{\partial t}\boldsymbol{\Pi}_{kSV}(t,\boldsymbol{x}) + \nabla\cdot\left(\boldsymbol{\Pi}_{kSV}(t,\boldsymbol{x})\,\boldsymbol{u}_{kSV}(t,\boldsymbol{x})\right) = \boldsymbol{Ge}_{kSV}(t,\boldsymbol{x}) \tag{8.7}$$

$$\boldsymbol{\Pi}_k(t,\boldsymbol{x}) = \left(\boldsymbol{\Pi}_{kQ}(t,\boldsymbol{x})\boldsymbol{\Pi}_{kSV}(t,\boldsymbol{x})\right);\ \boldsymbol{Ge}_k(t,\boldsymbol{x}) = \left(\boldsymbol{Ge}_{kQ}(t,\boldsymbol{x})\boldsymbol{Ge}_{kSV}(t,\boldsymbol{x})\right) \tag{8.8}$$

Economic equations on transactions $\boldsymbol{Tr}_k(t,\boldsymbol{x})$ (6.1-6.4) and their impulses $\boldsymbol{P}_k(t,\boldsymbol{x})$ (7.1-7.4) and on expected transactions $\boldsymbol{Et}_k(t,\boldsymbol{x})$ (8.1-8.4) and impulses $\boldsymbol{\Pi}_k(t,\boldsymbol{x})$ (8.5-8.8) establish a self consistent system of equations that model interactions between transactions and expectations. This system of equations describes the relations between transactions and expectations of separate economic agents and macroeconomic transactions and expectations.

## 5. Model interactions between transactions and expectations

Expectations $\boldsymbol{Ex}_k(t,\boldsymbol{x})$ those approve transactions $\boldsymbol{Tr}_k(t,\boldsymbol{x})$ may depend on numerous economic and financial variables, transactions and other expectations. Let's study simple model relations between transactions $\boldsymbol{Tr}_k(t)$ and expected transactions $\boldsymbol{Et}_k(t)$ of entire economics as functions of time $t$ only:

$$\boldsymbol{Tr}_k(t) = \int dx\ \ \boldsymbol{Tr}_k(t,\boldsymbol{x})\ ;\quad \boldsymbol{Et}_k(t) = \int dx\ \ \boldsymbol{Et}_k(t,\boldsymbol{x}) \tag{9.1}$$

Integrals in (9.1) are taken over economic domain (1.4) and due to equations (5.2.5) and (6.5) equations (6.1-6.4) on transactions $\boldsymbol{Tr}_k(t)$ and (8.1-8.4) on expected transactions $\boldsymbol{Et}_k(t)$ take form of ordinary differential equations:

$$\frac{d}{dt}\boldsymbol{Tr}_k(t) = \boldsymbol{F}_k(t) \;\; ; \;\; \frac{d}{dt}Q_k(t) = F_{kQ}(t) \; ; \; \frac{d}{dt}SV_k(t) = F_{kSV}(t) \quad (9.2)$$

$$\boldsymbol{F}_k(t) = \int d\boldsymbol{x} \; \boldsymbol{F}_k(t,\boldsymbol{x}) \;\; ; \;\; F_{kQ}(t) = \int d\boldsymbol{x} \; F_{kQ}(t,\boldsymbol{x}) \;\; ; \;\; F_{kSV}(t) = \int d\boldsymbol{x} F_{kSV}(t,\boldsymbol{x}) \quad (9.3)$$

$$\frac{d}{dt}\boldsymbol{Et}_k(t) = \boldsymbol{Fe}_k(t) \;\; ; \;\; \frac{d}{dt}Et_{kQ}(t) = Fe_{kQ}(t) \; ; \; \frac{d}{dt}Et_{kSV}(t) = Fe_{kSV}(t) \quad (9.4)$$

$$\boldsymbol{Fe}_k(t) = \int d\boldsymbol{x} \; \boldsymbol{Fe}_k(t,\boldsymbol{x}); \; Fe_{kQ}(t) = \int d\boldsymbol{x} \; Fe_{kQ}(t,\boldsymbol{x}); \; Fe_{kSV}(t) = \int d\boldsymbol{x} Fe_{kSV}(t,\boldsymbol{x}) \quad (9.5)$$

$$\boldsymbol{Et}_k(t) = \boldsymbol{Ex}_k(t)\boldsymbol{Tr}_k(t) \; ; \; Et_{kQ}(t) = Ex_{kQ}(t)Q_k(t) \; ; \; Et_{kSV}(t) = Ex_{kSV}(t)SV_k(t) \quad (9.6)$$

Equations (9.2-9.5) describe transactions $\boldsymbol{Tr}_k(t)$ (9.1) , $k=1,..K$ with assets under consideration performed by all agents of the entire economics under expectations $\boldsymbol{Ex}_k(t)=(Ex_{kQ}(t), Ex_{kSV}(t))$ determined by (9.6). Let's describe mutual action between disturbances of transactions and expectations in the linear approximation. To do that let's consider the system of equations (9.2-9.5) and assume that mean values $\boldsymbol{Tr}_{k0}(t)$ of transactions $\boldsymbol{Tr}_k(t)$ and mean values $\boldsymbol{Et}_{k0}(t)$ of expected transactions $\boldsymbol{Et}_k(t)$ are slow to compare with disturbances $\boldsymbol{tr}_k(t)$ of transactions and expected transactions $\boldsymbol{et}_k(t)$ and hence let's take them as constants:

$$\boldsymbol{Tr}_k(t) = \boldsymbol{Tr}_{k0}(1 + \boldsymbol{tr}_k(t)) \;\; ; \;\; \boldsymbol{Et}_k(t) = \boldsymbol{Et}_{k0}(1 + \boldsymbol{et}_k(t)) \quad (9.7)$$

Relations (9.7) present disturbances $\boldsymbol{tr}_k(t)$ and $\boldsymbol{et}_k(t)$ as dimensionless variables. Equations on disturbances take form:

$$Q_{k0} \frac{d}{dt}q_k(t) = f_{kq}(t) \; ; \; SV_{k0} \frac{d}{dt}sv_k(t) = f_{ksv}(t) \quad (9.8)$$

$$Et_{k0Q} \frac{d}{dt}et_{kq}(t) = fe_{kq}(t) \; ; \; Et_{k0SV} \frac{d}{dt}et_{ksv}(t) = fe_{ksv}(t) \quad (9.9)$$

$$Q_k(t) = Q_{k0}\big(1 + q_k(t)\big); \quad SV_k(t) = SV_{k0}\big(1 + sv_k(t)\big) \quad (9.10)$$

$$Et_{kQ}(t) = Et_{k0Q}\left(1 + et_{kq}(t)\right); \quad Et_{kSV}(t) = Et_{k0SV}\big(1 + et_{ksv}(t)\big) \quad (9.11)$$

$$\boldsymbol{tr}_k(t) = \big(q_k(t); sv_k(t)\big) \; ; \; \boldsymbol{et}_k(t) = \left(et_{kq}(t); et_{ksv}(t)\right) \quad (9.12)$$

$$\boldsymbol{F}_k(t) = \boldsymbol{F}_{k0}(t) + \boldsymbol{f}_k(t) \;\; ; \;\; \boldsymbol{Fe}_k(t) = \boldsymbol{Fe}_{k0}(t) + \boldsymbol{fe}_k(t) \quad (9.13)$$

$$\boldsymbol{f}_k(t) = \left(f_{kq}(t); f_{ksv}(t)\right) \;\; ; \boldsymbol{fe}_k(t) = \left(fe_{kq}(t); fe_{ksv}(t)\right) \quad (9.14)$$

Let's assume that factors $f_{kq}(t)$ and $f_{ksv}(t)$ in the right hand side of equations (9.8) depend on disturbances of expected transactions $et_{kq}(t)$ and $et_{ksv}(t)$ and factors $fe_{kq}(t)$ and $fe_{ksv}(t)$ in the right hand side of equations (9.9) depend on disturbances of transactions $q_k(t)$ and $sv_k(t)$. Thus for linear approximation let's take equations (9.8; 9.9) as:

$$Q_{k0} \frac{d}{dt}q_k(t) = a_{kq}Et_{k0Q}et_{kq}(t) \;\; ; \; SV_{k0}\frac{d}{dt}sv_k(t) = a_{ksv}Et_{k0SV} \; et_{ksv}(t) \quad (10.1)$$

$$Et_{k0Q} \frac{d}{dt}et_{kq}(t) = be_{kq}Q_{k0}q_k(t) \;\; ; \; Et_{k0SV}\frac{d}{dt}et_{ksv}(t) = be_{ksv}SV_{k0} \; sv_k(t) \quad (10.2)$$

For

$$\omega_{kq}^2 = -a_{kq}be_{kq} > 0 \; ; \; \omega_{ksv}^2 = - \, a_{ksv}be_{ksv} > 0 \quad (10.3)$$

equations (10.1; 10.2) on disturbances take form of equations for harmonic oscillators:

$$\left(\frac{d^2}{dt^2}+\omega_{kq}^2\right)q_k(t)=0\ ;\ \left(\frac{d^2}{dt^2}+\omega_{ksv}^2\right)sv_k(t)=0 \tag{10.4}$$

$$\left(\frac{d^2}{dt^2}+\omega_{kq}^2\right)et_{kq}(t)=0\ ;\ \left(\frac{d^2}{dt^2}+\omega_{ksv}^2\right)et_{ksv}(t)=0\ ;k=1,..K \tag{10.5}$$

Simple solutions of (10.4) for dimensionless disturbances $q_k(t)$ and $sv_k(t)$:

$$q_k(t)=c_{kq}sin\omega_{kq}t+\ d_{kq}cos\omega_{kq}t \tag{10.6}$$

$$sv_k(t)=c_{ksv}sin\omega_{ksv}t+d_{ksv}cos\omega_{ksv}t \tag{10.7}$$

$$c_{kq},d_{kq}\ ,c_{ksv},d_{ksv}\ll 1 \tag{10.8}$$

Relations (10.6-10.8) present simplest example of harmonic fluctuations of disturbances of trading volume $q_k(t)$ and trading value $sv_k(t)$ under $k=1,...K$ different expectation. As we show below (10.6-10.8) define price and return disturbances.

*5.1 Price fluctuations*

Disturbances of transactions under different expectations cause disturbances of price. Relations (2.2) and (9.1; 9.10) define price $p_k(t)$ of transactions $\boldsymbol{Tr}_k(t)$ performed under expectations of type $k$, $k=1,...K$ by all agents as:

$$SV_k(t)=p_k(t)Q_k(t) \tag{10.9}$$

Aggregate of transactions $\boldsymbol{Tr}_k(t)$ (9.1) over all types $k$, $k=1,...K$ of expectations define cumulative transactions $\boldsymbol{Tr}(t)$ in the entire economics with assets under consideration:

$$\boldsymbol{Tr}(t)=\sum_k\boldsymbol{Tr}_k(t)\ ;\ Q(t)=\sum_k Q_k(t)\ ;\ SV(t)=\sum_k SV_k(t) \tag{10.10}$$

Relations (9.1; 10.9) determine price $p(t)$ of cumulative transactions $\boldsymbol{Tr}(t)$ (10.10) as:

$$SV(t)=p(t)Q(t) \tag{10.11}$$

Relations (10.11) and (10.6-10.8) define price $p(t)$ as:

$$Q(t)=\sum_k Q_{k0}\big(1+q_k(t)\big)=Q_0\sum_k\lambda_k\big(1+q_k(t)\big) \tag{11.1}$$

$$SV(t)=\sum_k SV_{k0}\big(1+sv_k(t)\big)=SV_0\sum_k\mu_k\big(1+sv_k(t)\big) \tag{11.2}$$

$$Q_0=\sum_k Q_{k0}\ \ ;\ \lambda_k=\frac{Q_{k0}}{Q_0}\ ;\ SV_0=\sum_k SV_{k0}\ \ ;\ \mu_k=\frac{SV_{k0}}{SV_0} \tag{11.3}$$

$$\sum\lambda_k=\sum\mu_k=1 \tag{11.4}$$

$$p(t)=\frac{SV(t)}{Q(t)}=\frac{\sum_{k=1,..K}SV_k(t)}{\sum_{k=1,..K}Q_k(t)} \tag{11.5}$$

$$p_0=\frac{SV_0}{Q_0}=\frac{\sum_{k=1,..K}SV_{k0}}{\sum_{k=1,..K}Q_{k0}} \tag{11.6}$$

In linear approximation by disturbances $q_k(t)$ and $sv_k(t)$ price $p(t)$ (11.5) of assets under consideration can be presented as follows:

$$p(t)=p_0[1+\pi(t)]=p_0[1+\sum\mu_k sv_k(t)-\sum\lambda_k q_k(t)] \tag{11.7}$$

Relations (11.7) show that dimensionless price fluctuations $\pi(t)$ (11.8) depend on fluctuations of trade value $sv_k(t)$ and on fluctuations of trade volume $q_k(t)$:

$$\pi(t) = \sum \mu_k sv_k(t) - \sum \lambda_k q_k(t) \tag{11.8}$$

Relations (10.6-10.8) and (11.8) define price fluctuations with frequencies (10.3). Coefficients $\mu_k$ and $\lambda_k$ are determined by (11.3; 11.4) and describe impact of trade value $SV_{k0}$ and trade volume $Q_{k0}$. For each $k=1,..K$ let's present (10.9) as

$$SV_k(t) = SV_{k0}[1 + sv_k(t)] = p_{k0}[1 + \pi_k(t)]Q_{k0}[1 + q_k(t)] \tag{11.9}$$

Relations (11.9) in linear approximation by disturbances give:

$$SV_{k0} = p_{k0}Q_{k0} \quad ; \quad sv_k(t) = \pi_k(t) + q_k(t) \tag{11.10}$$

Substitution (11.10) into (11.8) gives

$$\pi(t) = \sum \mu_k \pi_k(t) + \sum(\mu_k - \lambda_k)q_k(t) \tag{11.11}$$

Relations (11.11) describe perturbations of price $\pi(t)$ as weighted sum of partial price disturbances $\pi_k(t)$ determined by different expectations for $k=1,..K$ and weighted sum of partial trade volume disturbances $q_k(t)$. Thus statistics of price disturbances $\pi(t)$ can depend on statistics of partial price disturbances $\pi_k(t)$ and on statistics of partial trade volume disturbances $q_k(t)$ for different expectations $k=1,..K$. One should take into account this issue while modeling statistical distribution of price fluctuations.

*5.2 Return fluctuations*

Price fluctuations (11.11) cause disturbances of return $r(t,d)$:

$$r(t,d) = \frac{p(t)}{p(t-d)} - 1 \tag{12.1}$$

To derive relations on return $r(t,d)$ for price $p(t)$ determined by (11.7; 11.11) let's introduce partial returns $r_k(t,d)$ (12.2) for price $p_k(t)$ as:

$$r_k(t,d) = \frac{p_k(t)}{p_k(t-d)} - 1 \tag{12.2}$$

and partial "returns" $w_k(t,d)$ (12.3) of trade volumes $Q_k(t)$ as

$$w_k(t,d) = \frac{Q_k(t)}{Q_k(t-d)} - 1 \tag{12.3}$$

Let's assume for simplicity that mean price $p_{0k}$ and trade volumes $Q_{k0}$ are constant during time term $d$ then (9.10; 11.7; 11.11) allow present (12.1; 12.3) as

$$r_k(t,d) = \frac{\pi_k(t)-\pi_k(t-d)}{1+\pi_k(t-d)} \quad ; \quad w_k(t,d) = \frac{q_k(t)-q_k(t-d)}{1+q_k(t-d)} \tag{12.4}$$

$$r(t,d) = \sum \mu_k \frac{1+\pi_k(t-d)}{1+\pi(t-d)} r_k(t,d) + \sum(\mu_k - \lambda_k)\frac{1+q_k(t-d)}{1+\pi(t-d)} w_k(t,d) \tag{12.5}$$

Let's define

$$\varepsilon_k(t-d) = \mu_k \frac{1+\pi_k(t-d)}{1+\pi(t-d)} \quad ; \quad \eta_k(t-d) = (\mu_k - \lambda_k)\frac{1+q_k(t-d)}{1+\pi(t-d)} \tag{12.6}$$

$$\sum[\varepsilon_k(t-d) + \eta_k(t-d)] = 1 \tag{12.7}$$

$$r(t,d) = \sum \varepsilon_k(t-d) r_k(t,d) + \sum \eta_k(t-d) w_k(t,d) \tag{12.8}$$

Relations (12.4-12.8) describe dependence of return (12.1) on partial returns $r_k(t,d)$ and on "returns" $w_k(t,d)$ of trade volumes $Q_k(t)$ (12.2-12.4).

Above relations describe price $p(t)$ (11.7) and price disturbances $\pi(t)$ (11.11) as weighted sum of partial price disturbances $\pi_k(t)$ and partial trade volume disturbances $q_k(t)$ determined by numerous expectations for $k=1,..K$. Relations (12.8) present return $r(t,d)$ (12.1) as weighted sum of partial returns $r_k(t,d)$ (12.2) and trade volume "returns" $w_k(t,d)$ (12.3). Sum for coefficients $\mu_k$ and $\mu_k$-$\lambda_k$ for price $p(t)$ (11.11) and $\varepsilon_k(t)$ and $\eta_k(t)$ for return $r(t,d)$ (12.8) equals unit but (11.11) and (12.8) can't be treated as certain averaging as some coefficients $\mu_k$-$\lambda_k$ and $\eta_k(t)$ should be negative. If mean price $p_{k0}$ (11.10) is constant for all expectations $k=1,..K$ and

$$p_0 = p_{k0} \;\; ; \;\; k = 1,..K \tag{13.1}$$

then it is easy to show that

$$\lambda_k = \mu_k \;\; ; \;\; \eta_k(t) = 0 \; for \; all \; k = 1,..K \tag{13.2}$$

and relations (11.11; 12.8) take simple form

$$\pi(t) = \sum \mu_k \pi_k(t) \tag{13.3}$$

$$r(t,d) = \sum \varepsilon_k(t-d) r_k(t,d) = \sum \mu_k \frac{\pi_k(t)-\pi_k(t-d)}{1+\pi(t-d)} \tag{13.4}$$

We propose that relations (13.1) on price $p_{k0}$ may fail for transactions driven by different kinds of expectations. Expectations are key factor for market competition and different expectations may cause different mean partial prices. That should cause more complex representation of price (11.11) and return (12.8) disturbances as well as impact on volatility and statistic distributions of price and return disturbances.

If one takes into account possible linear price trend during time term $d$ and π(t) as (11.11)

$$p(t) = p_0\left(1 + \alpha t + \pi(t)\right)$$

then for return $r(t,d)$ (12.1) obtain (see Appendix):

$$r(t,d) = \frac{\alpha d}{1 + \alpha(t-d) + \pi(t-d)} + \sum \varepsilon_{k2}(t-d) \; r_k(t,d) + \eta_{k2}(t-d) w_k(t,d)$$

## 6. Conclusion

Economic space permits describe transactions and expectations as functions of risk coordinates and derive economic equations on transactions and expectations. We propose that agents make transactions under different kinds of expectations. As example we study simple model interactions between transactions and expectations and describe relations between transactions and expectations in a closed form. For this simple example we obtain that trade volume and price disturbances can be described by fluctuations with numerous frequencies (10.3). Different kinds of expectations define partial transactions with different prices and trade volume disturbances. We derive representation of cumulative price disturbances determined by all expectations as weighted sum of partial price and trade volume disturbances. Description of price fluctuations allows model evolution of return and we present cumulative return defined by all expectations as weighted sum of partial return and trade volume "return" (12.3).

Such representation of price and return for very simple model indicates that their statistical distributions can depend on statistical properties of trade volume disturbances and trade volume "return" respectively. Thus description of price and return volatility and their statistical distributions should take into account possible dependence on statistical properties of trade volume disturbances and trade volume "return". These relations should be studied further.

## Appendix

### Price and return with linear price trend

To take into account possible linear price trend during time term $d$ let's take

$$Q_k(t) = Q_{k0}\big(1 + \gamma_k t + q_k(t)\big) \quad ; \quad Q_0(t) = \sum Q_{k0}\,(t) = Q_0\big(1 + \gamma t + q(t)\big)$$

$$SV_k(t) = SV_{k0}\big(1 + \beta_k t + sv_k(t)\big) \; ; \quad SV_0(t) = \sum SV_{k0}\,(t) = SV_0\big(1 + \beta t + sv(t)\big)$$

$$p_k(t) = p_{k0}(1 + \alpha_k t + \pi_k(t)) \quad ; \quad p(t) = p_0\big(1 + \alpha t + \pi(t)\big)$$

$$Q_0 = \sum Q_{k0} \;\; ; \;\; Q_0\gamma = \sum Q_{k0}\gamma_k \quad ; \quad Q_0 q(t) = \sum Q_{k0}\, q_k(t)$$

$$SV_0 = \sum SV_{k0} \;\; ; \;\; SV_0\beta = \sum SV_{k0}\beta_k \quad ; \quad SV_0 sv(t) = \sum SV_{k0}\, sv_k(t)$$

$$\mu_k = \frac{SV_{k0}}{SV_0} \;\; ; \;\; \lambda_k = \frac{Q_{k0}}{Q_0} \;\; ; \;\; \alpha_k = \mu_k\beta_k - \lambda_k\gamma_k \;\; ; \;\; \alpha = \sum(\mu_k\beta_k - \lambda_k\gamma_k) = \sum \alpha_k$$

$$\beta = \sum \frac{SV_{k0}}{SV_0}\beta_k = \sum \mu_k\beta_k \quad ; \quad \gamma = \sum \frac{Q_{k0}}{Q_0}\gamma_k = \sum \lambda_k\gamma_k$$

$$p(t) = p_0\big(1 + \alpha t + \pi(t)\big) = \frac{SV_0\big(1 + \beta t + sv(t)\big)}{Q_0\big(1 + \gamma t + q(t)\big)} = \frac{SV_0}{Q_0}\,\big(1 + (\beta - \gamma)t + sv(t) - q(t)\big)$$

$$p_0 = \frac{SV_0}{Q_0} \quad ; \quad \alpha = (\beta - \gamma) \quad ; \quad \pi(t) = sv(t) - q(t)$$

Due to (11.10)

$$p(t) = p_0\,\big(1 + \alpha t + \pi(t)\big) \quad ; \quad \pi(t) = \sum \mu_k\pi_k(t) + \sum(\mu_k - \lambda_k)q_k(t)$$

For return *r(t,d)* (12.1) obtain:

$$r(t,d) = r_1(t,d) + r_2(t,d)$$

$$r_1(t,d) = \frac{p_0\alpha d}{p(t-d)} \quad ; \quad r_2(t,d) = \frac{p_0 a[\pi(t) - \pi(t-d)]}{p(t-d)}$$

For $r_k(t,d)$ (12.2) and $w_k(t,d)$ (12.3) obtain:

$$r_2(t,d) = \sum \varepsilon_{k2}(t-d)\; r_k(t,d) + \eta_{k2}(t-d)w_k(t,d)$$

$$\varepsilon_{k2}(t-d) = \mu_k \frac{1 + \pi_k(t-d)}{1 + \alpha(t-d) + \pi(t-d)}$$

$$\eta_{k2}(t-d) = (\mu_k - \lambda_k)\frac{1 + q_k(t-d)}{1 + \alpha(t-d) + \pi(t-d)}$$

$$r(t,d) = \frac{\alpha d}{1 + \alpha(t-d) + \pi(t-d)} + \sum \varepsilon_{k2}(t-d)\; r_k(t,d) + \eta_{k2}(t-d)w_k(t,d)$$